\documentclass[pra,twocolumn,amsmath,amssymb,byrevtex]{revtex4}

\usepackage{bm,amssymb,amsmath,graphicx,color}

\newtheorem{Thm}{Theorem}\newtheorem{Prop}{Proposition}\newtheorem{Def}{Definition}\newcommand{\tr}{\mathop{\mathrm{tr}}\nolimits}\newcommand{\HA}{\mathop{\mathcal{H}}\nolimits}\newcommand{\LA}{\mathop{\mathcal{L}}\nolimits}\newcommand{\SA}{\mathop{\mathcal{S}}\nolimits}\newcommand{\R}{\mathop{\mathbb{R}}\nolimits}\newcommand{\I}{\mathop{\mathbb{I}}\nolimits}\newcommand{\E}{\mathop{\mathcal{E}}\nolimits}

\newcommand{\bra}[1]{\langle #1 |}\newcommand{\ket}[1]{| #1 \rangle}
\newcommand{\ketbra}[2]{| #1 \rangle \langle #2 |}
\begin{document}

\title{Optimal State Discrimination in General Probabilistic Theories}
\author{Gen Kimura ${}^{[a]}$}
\email{gen-kimura[at]aist.go.jp}
\author{Takayuki Miyadera ${}^{[a]}$}
\email{miyadera-takayuki[at]aist.go.jp}
\author{Hideki Imai ${}^{[a],[b]}$}
\affiliation{[a] Research Center for Information Security (RCIS),
National Institute of Advanced Industrial
Science and Technology (AIST). 
Daibiru building 1003,
Sotokanda, Chiyoda-ku, Tokyo, 101-0021, Japan \\ 
[b] Graduate School of Science and Engineering,
Chuo University.
1-13-27 Kasuga, Bunkyo-ku, Tokyo 112-8551, Japan
}

\begin{abstract}
We investigate a state discrimination problem in operationally the most general framework to use a probability, including both classical, quantum theories, and more. 
In this wide framework, introducing closely related family of ensembles (which we call a {\it Helstrom family of ensembles}) with the problem, we provide a geometrical method to find an optimal measurement for state discrimination by means of Bayesian strategy. 
We illustrate our method in $2$-level quantum systems and in a probabilistic model with square-state space to reproduce e.g., the optimal success probabilities for binary state discrimination and $N$ numbers of symmetric quantum states. 
The existences of families of  ensembles in binary cases are shown both in classical and quantum theories in any generic cases.  
\end{abstract}
\maketitle

\section{Introduction}

Among many attempts to understand quantum theory axiomatically,
an operationally natural approach has attracted increasing attention
in the recent development of quantum information
theory \cite{ref:H,ref:F,ref:B1,ref:CBH,ref:D}.
By constructing a general framework of theories to include not only classical and quantum theories but also more general theories,  
one can reconsider the nature of quantum theory from outside, preferably with the operational and informational point of view.
This also enables us to prepare for a (possible) post-quantum theory in the future. For instance, it is important to find conditions to achieve a secure key distribution in a general framework
\cite{ref:BHK}.
Among others, the convexity or operational approach \cite{ref:G}, or recently referred as ``general (or generic) probabilistic theories (or models)" \cite{ref:Clone,ref:Tel}, is considered to provide operationally the most general theory for probability. 
Of course, both classical probability theory and quantum theory are included as typical examples of general probabilistic theories, but it is known that there exist other possible physical models for probability (See an example in Sec. IV B). 

Although this approach has relatively long history \cite{ref:Old,ref:HolevoSM}, there are still many fundamental problems especially from the applicational and informational points of view to be left open. 
This may not be surprising if one recalls that quantum information theory has given new insights and provided attractive problems on the foundation and application of quantum mechanics.
One of them is a state discrimination problem.
The problem asks how well a given ensemble of states is distinguishable.
It has been one of the most important questions in quantum information theory, and there are various
formulations of the problem depending on measures to characterize the quality of discrimination \cite{Helstrombook,IDP,Yuen,Fuchs}.
The property that there is no measurement perfectly distinguishes non-orthogonal pure states plays an essential role in the various protocols such as quantum key distribution \cite{BB84}, and is often considered as the most remarkable feature of quantum theory.
On the other hand, in the context of general probabilistic theories, the property can characterize the nature of classical theory. 
Indeed, it is known that a general probabilistic theory is a classical theory if and only if all the pure states can be perfectly discriminated in a single measurement \cite{ref:Clone}.
\par
In this paper, we discuss an optimal state discrimination problem in general probabilistic theories by means of Bayesian strategy.  
While the existence of Bayes optimal measurements has been discussed in general setting \cite{ref:O}, we provide a geometrical method to find such optimal measurement and optimal success probability. 
Our figure of merit is the optimal success probability, in discriminating $N$ numbers of states under a given prior distribution.
We introduce a useful family of ensembles, which we call a {\it Helstrom family of ensembles}, in any general probabilistic theories, which generalizes a family of ensembles used in \cite{ref:Hwang} in $2$-level quantum systems for binary state discrimination, and show that the family enable us to obtain optimal measurements by means of Bayesian strategy. 
This method reveals that a certain geometrical relation between state space and the convex subset generated by states which we want to distinguish is crucial for the problem of state discrimination: In the case of uniform prior distribution, what one has to do is to find as large convex subset (composed of Helstrom family of ensembles) as possible in state space which is reverse homothetic to the convex subset generated by states under consideration. 
The existences of the Helstrom families for $N=2$ which again have a simple geometrical interpretation are shown in both classical and quantum systems in generic cases.
Some other works on the problem in quantum theory are related with our purpose; 
The no-signaling condition was used in deriving the optimal success probability \cite{ref:Hwang} between two states in $2$-level quantum systems, a bound of the optimal success probability \cite{ref:UnambiSD}
and a maximal confidence \cite{ref:CAB} among several non-orthogonal states in general quantum systems. 
In particular, we discuss the relation between our method and the one used in \cite{ref:Hwang}, and show that our method generalizes the results in \cite{ref:Hwang} to general probabilistic theories.

The paper is organized as follows. 
In Sec.~\ref{sec:review}, we give a brief review of general probabilistic theories. 
In Sec.~\ref{sec:1}, we introduce a {\it Helstrom family of ensembles} and show the relation with an optimal measurement in state discrimination problem (Propositions \ref{Prop:BDD}, \ref{Prop:SC}, Theorem \ref{thm:HE2}).
We also prove the existences of the families of ensembles for $N=2$ in classical and quantum systems in generic cases (Theorems \ref{thm:QHE}, \ref{thm:CHE}). 
In Sec.~\ref{sec:ex}, we illustrate our method in $2$-level quantum systems, and reproduce the optimal success probabilities for binary state discrimination and $N$ numbers of symmetric quantum states. 
As an example of neither classical nor quantum theories, we introduce a general probabilistic model with square-state space.
Our method is also applied to this model to exemplify its usability.  
In Sec.~\ref{ref:2}, we summarize our results. 

\section{Brief Review of General Probabilistic Theories}\label{sec:review}
In order to overview general probabilistic theories as the operationally most general theories of probability, 
let us start from a very primitive consideration of physical theories where a probability plays a fundamental role. 
In such a theory, a particular rule (like Borel rule in Quantum mechanics) to obtain a probability for some output when measuring an observable ${\bm o}$ under a state $s$ should be provided. 
Therefore, states and observables are two fundamental ingredients with an appropriate physical law to obtain probabilities in general probabilistic theories.  
Let us denote the set of states by ${\cal S}$.  
In a simplified view, an $N$-valued observable ${\bm o}$ \footnote{In this paper, we deal with only finite discrete observables with finitely many outputs, since it is enough to consider for our purpose to discriminate $N$ numbers of states. 
Note that it is straightforward to formalize general observables with measure theoretic language. } can be considered as an $N$ numbers of maps $o_i $ on a state space $\SA$ so that $o_i(s) \in [0,1]$ provides a probability to obtain $i$th output when measuring this observable under a state $s \in \SA$. 
It is operationally natural to assume that if one can prepare states $s \in \SA$ and $t \in \SA$, then there exists a probabilistic-mixture state $<\lambda,s,t> \in \SA$ for any $\lambda \in [0,1]$ which represents an ensemble of preparing state $s$ with probability $\lambda$ and state $t$ with probability $1-\lambda$. 
Furthermore, it is natural to assume the so-called separating condition for states; namely, two states $s_1$ and $s_2$ should be identified when there are no observables to statistically distinguish them. 
Then, it has been shown \cite{ref:G,ref:O} that without loss of generality, the state space  $\SA$ is embedded into a convex (sub)set in a real vector space $V$ such that a probabilistic-mixture state is given by a convex combination $<\lambda,s,t> = \lambda s + (1-\lambda) t$ \footnote{A subset $C$ in a real vector space $V$ is called convex if $\lambda s + (1-\lambda )t \in C$ for any $s,t \in C$ and $\lambda \in [0,1]$. }. 
Hence, hereafter the state space $\SA$ is assumed to be convex set in a real vector space $V$ with the above mentioned interpretation. 
An extreme point \footnote{If $s \in C$ does not have a nontrivial convex combination in $C$, i.e., if $s = \lambda t + (1-\lambda)u $ for some $t,u \in C$ and $\lambda \in (0,1)$ implies $s = t = u$, then $s$ is called an extreme point.} of a state space $\SA$ is called a pure state, otherwise a mixed state. 
Physically, a pure state is a state which cannot be prepared as an ensembles of different states.  
From the preparational point of view for state $<\lambda,s,t>=\lambda s + (1-\lambda) t$, each maps $o_i$ of an observable ${\bm o}$ should be an affine functional: $o_i(\lambda s + (1-\lambda) t) = \lambda o_i(s) + (1-\lambda) o_i(t)$, since the right hand side is a sum of probabilities to obtain $i$th outputs for exclusive events of states $s$ and $t$ with probability $\lambda$ and $1-\lambda$, while $ o_i(s), o_i(t) $ are conditional probabilities to obtain $i$th output conditioned that the states are $s$  and $t$, respectively.    
An effect $e$ on $\SA$ is an affine functional from $\SA$ to $[0,1]$.   
There are two trivial effects, unit effect $u$ and zero effect $0$, defined by $u(s) = 1, 0(s) = 0 \ $ for all $s \in \SA$. 
With this language, an $N$-valued observable ${\bm o} $ is a set of effects $o_i \ (i=1,\ldots,N)$ satisfying $\sum_{i=1}^N o_i = u$, meaning that $o_i(s)$ is the probability to obtain the $i$th output when measuring the observable ${\bm o}$ in the state $s$. 
We denote by ${\cal E}$ and ${\cal O}_N$ the sets of all the effects and $N$-valued observables, respectively. 
While the output of an observable can be not only from real numbers but also any symbols, like ``head" or ``tail", hereafter we often identify them with $\{1,\ldots,N\}$. 
Physically natural topology on $\SA$ is given by the (weakest) topology so that all the effects are continuous. 
Without loss of generality \cite{ref:ExistHE}, $\SA$ is assumed to be compact with respect to this topology.   
Typical examples of the general probabilistic theories will be classical and quantum systems. 
For simplicity, the classical and quantum systems we consider in this paper will be finite systems: 
\bigskip 

[Example 1: Classical Systems]  Finite classical system is described by a finite probability theory.  
Let $\Omega=\{\omega_1,\ldots,\omega_d\}$ be a finite sample space. 
A state is a probability distribution $p = (p_1,\ldots,p_d)$, meaning that the probability to observe $\omega_i $ is $p_i$. 
Therefore, the state space is $\SA_{\mathrm{cl}} = \{p = (p_1,\ldots,p_d) \in \R^d  \ | \  p_i \ge 0, \ \sum_i p_i = 1 \} \subset \R^d$, and forms a (standard) simplex \footnote{A convex polytope $C= \mathrm{conv}_{i=1,\ldots,N}\{c_i\} = \{\sum_{i=1}^N p_i c_i \ | \ p_i \ge 0, \sum_i p_i = 1\} \subset V$ with $N$ numbers of extreme points $c_i \in V$ is called a simplex if any element $c \in C$ has the unique convex combinations with respect to $c_i$s. 
Equivalently, $C$ is called a simplex iff the affine dimension of $C$ is $N-1$.}. 
The set of extreme points is $\{p^{(i)}\}_{i=1}^d$ where $p^{(i)}_j = \delta_{ij}$.  
An effect $e$ is given by a random variable $f: \Omega \to [0,1]$ such that $e(p) = \sum_i f(\omega_i) p_i$ ($0 \le f(i) \le 1$). 
\bigskip 

[Example 2: Quantum Systems] $d$-level quantum system is described by an $d$ dimensional Hilbert space $\HA$.  
A state is described by a density operator $\rho$, an Hermitian positive operator on $\HA$ with unit trace, and the state space is given by $\SA_{\mathrm{qu}} = \{\rho \in \LA_H(\HA) \ | \ \rho \ge 0, \ \tr \rho= 1 \} \subset \LA_H(\HA)$; here real vector space $\LA_H(\HA)$ is the set of all Hermitian operator on $\HA$. 
A pure state is a one dimensional projection operator onto a unit vector $\psi \in \HA$, written as $\rho = \ketbra{\psi}{\psi}$ in Dirac notation.  	
An effect $e$ is described \cite{ref:O,ref:O2} by a positive operator $B$ such that $0 \le B \le \I$ through $e(\rho) = \tr (B \rho)$, which is called an element of positive-operator-valued measure (POVM) \footnote{Note that the observable here will be the so-called POVM (positive-operator-valued measure), and therefore what is usually called observable in the standard textbook of quantum mechanics which is characterized by an Hermitian operator is a special observable in this paper. }.  

In the following, we assume that all observables $\{o_i\}_{i=1}^N$ composed of effects $o_i$ satisfying $\sum_i o_i = u$ are in principle measurable. 
Then, only the structure of state space characterizes the general probabilistic theories. 
Roughly speaking, for each (compact) convex set one can consider the corresponding general probabilistic model. 
When we consider a composition of state spaces, the so-called no-signaling condition is usually required to keep the causality.

We refer \cite{ref:G,ref:HolevoSM} for the details of general probabilistic theories and \cite{ref:Clone} where generalized No-broadcasting and No-cloning theorems have been shown in general probabilistic theories.

\section{Helstrom Family of ensembles in General Probabilistic Theories}\label{sec:1}

As a state discrimination is one of the central problems in quantum information theory, we consider a problem to discriminate states in general probabilistic theories by means of Bayesian strategy. 
Suppose Alice is given a state chosen from $\{s_i \in \SA\}_{i=1}^{N}$ with a prior probability distribution $\{p_i \in \R \}_{i=1}^{N}$ ($p_i \ge0, \sum_i p_i = 1$), and her goal is to guess the state. 
She wants to find an optimal measurement to maximize the success probability.  
Without loss of generality, it is sufficient to consider an $N$-valued observable ${\bf E} = \{e_i\}_{i=1}^N \in {\cal O}_N$ from which she decides the state was in $s_i $ when obtaining the output $i$.  
Then, the success probability is  
\begin{equation}\label{eq:SP}
P_S({\bf E}) = \sum_{i=1}^N p_i e_i(s_i). 
\end{equation}
The optimal success probability $P_S$ is given by optimizing $P_S({\bf E})$ among all the $N$-valued observables: 
\begin{equation}\label{eq:HB}
P_S = \sup_{{\bf E} \in {\cal O}_N} P_S({\bf E}). 
\end{equation}
For a binary discrimination ($N=2$), it can be written as 
\begin{equation}\label{eq:HB2}
P_S  =  p_2 + \sup_{e \in \E} [p_1 e(s_1) - p_2 e(s_2)],
\end{equation}
where in the final expression we have used $e_1+ e_2 = u$. 
This problem is well investigated in quantum mechanics, and the optimal success probability to discriminate two distinct density operators $\rho_1,\rho_2$ with a prior distribution $p_1,p_2$ is given by 
\begin{eqnarray}\label{eq:OS}
P^{(Q)}_{S} = p_2 + \sup_{ 0 \le E \le \I} \tr [ E (p_1 \rho_1  - p_2 \rho_2)] \nonumber \\ 
= \frac{1}{2} (1 + ||p_1 \rho_1 - p_2 \rho_2||_1).
\end{eqnarray}
Here, the norm is a trace norm defined by $||A||_1 := \tr |A| = \tr \sqrt{A^\dagger A}$. 
Since this bound is sometimes referred as the Helstrom bound, let us call $P_S$ \eqref{eq:HB} also the Helstrom bound for any $N$ and for any general probabilistic theories.

In order to obtain the Helstrom bound in general probabilistic theories, we shall introduce a family of ensembles which is later shown to be closely related to the optimizing problem in Eq~\eqref{eq:HB}. 
In the following, we assume that a prior probability distribution satisfy $p_i \neq 0,1$ removing trivial cases: 
\begin{Def} 
Given $N$ distinct states $\{s_i \in \SA\}_{i=1}^N$ and a prior probability distribution $\{p_i\}_{i=1}^N$, 
we call a family of $N$-numbers of ensembles $\{\tilde{p}_i,s_i; 1 - \tilde{p}_i, t_i\} \ (i=1,\ldots,N)$ 
a ``weak Helstrom family of ensembles" (or simply a ``weak Helstrom family") for states $\{s_i\}$ and a probability $\{p_i\}$ if there exist $N$-numbers of binary probability distributions $\{\tilde{p}_i, 1- \tilde{p}_i \} \ (0< \tilde{p}_i \le 1)$ and $N$-numbers of states $\{t_i \in \SA\}_{i=1}^N$ satisfying 
\begin{eqnarray}
&\mathrm{(i)}& \frac{p_i}{\tilde{p}_i} = \frac{p_j}{\tilde{p}_j}\le 1, \label{eq:Hratio}\\
&\mathrm{(ii)}& \tilde{p_i} s_i + (1-\tilde{p_i}) t_i = \tilde{p_j} s_j + (1-\tilde{p_j}) t_j, \label{eq:HE}
\end{eqnarray}
for any $i,j=1,\ldots,N$ 
\end{Def}
\noindent 
Note that condition \eqref{eq:HE} means that $N$ ensembles $\{\tilde{p}_i,s_i; 1 - \tilde{p}_i, t_i\}$ are statistically equivalent (among observables). 
Therefore, a weak Helstrom family is a family of statistically equivalent ensembles which are mixtures of states $\{s_i\}$ and $\{t_i\}$ with weights $\tilde{p}_i$ and $1-\tilde{p}_i$ satisfying condition \eqref{eq:Hratio}. 
We call $t_i$ a {\it conjugate state} to $s_i$. 
The probabilistic-mixture state determined by $N$ ensembles $\{\tilde{p}_i,s_i; 1 - \tilde{p}_i, t_i\}$ with condition \eqref{eq:HE} is called a {\it reference state} and is denoted by $s$:
\begin{equation}
s := \tilde{p_i} s_i + (1-\tilde{p_i}) t_i \ (\forall i=1,\ldots,N). 
\end{equation}
We call the ratio $p \le 1$ a {\it Helstrom ratio} defined by 
\begin{equation}
p := \frac{p_i}{\tilde{p}_i} \ (\forall i = 1,\ldots,N),
\end{equation} 
which turns out to play an important role in an optimal state discrimination. 
We call a weak Helstrom family a {\it trivial (resp. nontrivial) family} when $p=1$ (resp.  $p<1$). 

Note that a weak Helstrom family always exists for any distinct states $\{s_i\}$ and a prior probability distribution $\{p_i\}$. 
For instance, it is easy to see that $\tilde{p_i} = p_i \ (p=1)$ and $t_i = \frac{1}{1-p_i}(\sum_{j\neq i} p_j s_j)$ gives a weak Helstrom family of ensembles with a reference state $s = \sum_i p_i s_i$, although it is a trivial family. (See later examples for nontrivial families.) 
Moreover, if $\{\tilde{p}_i,s_i; 1 - \tilde{p}_i, t_i\} \ (i=1,\ldots,N)$ is a weak Helstrom family with a Helstrom ratio $p < 1$ and a reference state $s$, then for any $p< p^\prime \le 1$,  one can construct another weak Helstrom family with a Helstrom ratio $p^\prime$.  
Indeed, since $0 \le \frac{1-\tilde{p}_i}{1-\tilde{p_i}^\prime} < 1$ for $\tilde{p}_i^\prime = \frac{pi}{p^\prime} \ (<1)$, one can take conjugate states as $t_i^\prime := \frac{1-\tilde{p}_i}{1-\tilde{p_i}^\prime} t_i + (1- \frac{1-\tilde{p}_i}{1-\tilde{p_i}^\prime}) s_i  $.
Then it is easy to see that the family of $\{s_i,\tilde{p}_i^\prime; t_i^\prime, 1-\tilde{p}_i^\prime\}$ is a weak Helstrom family with a Helstrom ratio $p^\prime$ and the same reference state $s$. 

\begin{figure} 
\includegraphics[height=0.25 \textwidth]{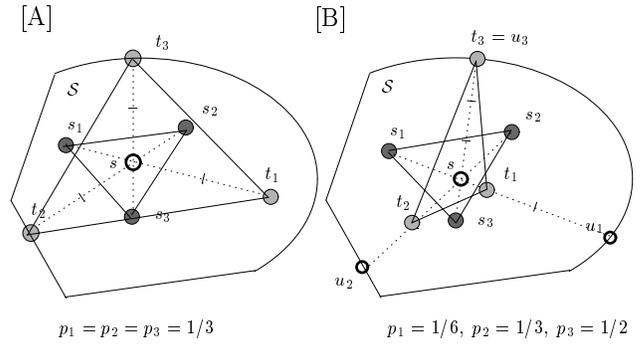}
\caption{Let $\SA$ be a convex set in $\R^2$ as depicted in the figures. 
For three distinct states $\{s_1,s_2,s_3\}$ in  $\SA$, non-trivial weak Helstrom families are illustrated [A]  for the uniform distribution and [B] for $p_1=1/6,p_2=1/3,p_3=1/2$, where Helstrom ratios are [A] $p= 1/3q =1/2 \ (q= \tilde{p_i} = 2/3)$ and [B]  $p= 2/3$. In [A], two polytopes (triangles) generated by $\{s_i\}$ and $\{t_i\}$ are reverse homothetic to one another with the similarity point $s$, while in [B], these polytopes are distorted homothetic depending on the prior distribution. }\label{fig:WHE}
\end{figure}

Let us explain a geometrical meaning of a weak Helstrom family of ensembles which makes it easier to find it. 
First we explain this for the most interesting cases in the context of state discrimination, i.e., those with the uniform probability distribution $p_i = 1/N$. 
In these cases, condition \eqref{eq:Hratio} tells that $\tilde{p_i}$ should give the same weights $q := \tilde{p_i} = \frac{1}{Np}$, and condition \eqref{eq:HE} geometrically means that all $t_i$ should located in $\SA$ such that all $s_i$ and $t_i$ have the common interior point $s$ with the same ratio $q$. 
Global picture for this is that one has to find $t_i$ so that the polytopes $X=\mathrm{conv}_{i=1,\ldots,N}[t_i]$ as a subset of $\SA$ and $Y=\mathrm{conv}_{i=1,\ldots,N}[s_i]$ posses the internal homothetic center $s$ in $\SA$ so that the polytopes $X$ and $Y$ are geometrically similar to one another with the similarity ratio $\frac{q}{1-q}$. 
Fig.~\ref{fig:WHE} [A] illustrates an example for $N=3$ with the uniform distribution. 
One immediately recognizes the reverse homothethy between two polytopes (triangles) generated by $\{s_i\}$ and $\{t_i\}$ with the internal homothetic center $s$. 
As is later shown, it is preferable to find a weak Helstrom family with smaller $p$ (and hence larger $q$) as much as possible. 
Therefore, if one knows the global image of state space $\SA$, then finding as large polygon as possible in $\SA$ which is reverse homothetic to the polygon generated by $\{s_i\}$ will provide you a good weak Helstrom family. 
Another simple algorithm to find a weak Helstrom family is the following: First choose freely a reference state $s$, and making lines from each $s_i$ passing through $s$ to the point in $\SA$ with which $s$ is the interior point with the common ratio $q$ and $1-q$ (See Fig.~\ref{fig:WHE} [A] for $N=3$). 
Then, with conjugate states as end-points of these lines, one obtains a weak Helstrom family $\{q,s_i; 1-q,t_i\}$ with a Helstrom ratio $p= \frac{1}{qN}$.  

With general prior probability distribution $\{p_i\}$, an algorithm to find a (possibly nontrivial) weak Helstrom family as small $p$ as possible is as follows: 
Take a reference state in $\SA$, e.g., $s = \sum_i p_i s_i$.  
Extend a line from each $s_i (i=1,\ldots,N)$ passing through $s$ until the line reaches the boundary of $\SA$.  
Let $u_i$ be such states on the boundary and let $0 \le q_i \le 1$ be the ratio so that $s= q_i s_i + (1-q_i)u_i$.
Then, conjugate states $t_i$ on each line satisfying $s = \tilde{p_i} s_i + (1-\tilde{p}_i)t_i$ with $\tilde{p}_i := \frac{p_i q_{i_0}}{p_{i_0}}$ where $i_0 := \mathrm{argmax}_{i=1,\ldots,N}[\frac{p_i}{q_i}]$, give a (nontrivial) weak Helstrom family of ensembles with a Helstrom ratio $p = \frac{p_{i_0}}{q_{i_0}}$. 
Notice that for general cases, the similarity between two polytopes generated by $\{s_i\}$ and $\{t_i\}$ is distorted. (See Fig.~\ref{fig:WHE} [B] for $N=3$.)    
   
In the following, we show that a weak Helstrom family of ensembles is closely related to an optimal state discrimination strategy, and provide a geometrical method to obtain the Helstrom bound $P_S$ and an optimal measurement in any general probabilistic theories. 

Let us again consider a state discrimination problem from $\{s_i \in \SA\}_{i=1}^N$ with a prior distribution $\{p_i\}_{i=1}^N$.
Let ${\bf E} = \{e_i \}_{i=1}^N$ be any $N$-valued observable from which Alice decides the state be in $s_i$ if she observes an output $i$. 
Suppose that we have a weak Helstrom family $\{\tilde{p}_i,s_i; 1 - \tilde{p}_i, t_i \} \ (i=1,\ldots,N)$ with the reference state $s = \tilde{p_i} s_i + (1-\tilde{p_i}) t_i \ (i=1,\ldots,N)$ and a Helstrom ratio $p = \frac{p_i}{\tilde{p}_i}$.   
Then, using $u = \sum_i e_i$, affinity of $e_i$ and Eq.~\eqref{eq:SP}, it follows 
\begin{eqnarray}\label{eq:keyineq}
1 = u(s)=\sum_i e_i (s) =  \sum_i e_i (\tilde{p_i} s_i + (1-\tilde{p_i}) t_i) \nonumber \\
 =  \frac{1}{p} \sum_i p_i e_i (s_i) +  \sum_i (1-\tilde{p}_i ) e_i(t_i) \nonumber \\
=  \frac{1}{p} P_S({\bf E}) +  \sum_i (1-\tilde{p}_i ) e_i(t_i).
\end{eqnarray}
Since $\sum_i (1-\tilde{p}_i ) e(t_i) \ge 0$, we obtain 
\begin{equation}
P_S({\bf E}) \le p
\end{equation}
which holds for any observables ${\bf E}$. Thus we have proved the following proposition.   
\begin{Prop}\label{Prop:BDD} Let $\{\tilde{p}_i,s_i; 1 - \tilde{p}_i, t_i\} \ (i=1,\ldots,N)$ be a weak Helstrom family of ensembles with a Helstrom ratio $p = \frac{p_i}{\tilde{p}_i} $.  
Then, we have a bound for the Helstrom bound $P_S \le p$. 
\end{Prop}
This means that, once we find a weak Helstrom family of ensembles, a bound of the Helstrom bound is automatically obtained. 
A trivial weak Helstrom family gives a trivial condition $P_S \le p=1$, which is the reason we called it trivial. 
Examples of nontrivial weak Helstrom families are given in Fig.~\ref{fig:WHE}, where [A] $P_S \le p= 1/2$ and [B] $P_S \le p= 2/3$.
Namely, the optimal success probability in this general probabilistic model is at most $1/2$ and $2/3$ for [A] $p_1=p_2=p_3=1/3$ and [B] $p_1=1/6,\ p_2=1/3,\ p_3 = 1/2$, respectively.   
   
Moreover, Proposition \ref{Prop:BDD} leads us to a useful notion of Helstrom family of ensembles defined as follows:  
\begin{Def}
Let $\{\tilde{p}_i,s_i; 1 - \tilde{p}_i, t_i\} \ (i=1,\ldots,N)$ be a weak Helstrom family of ensembles for $N$ distinct states $\{s_i\}$ and a prior probability distributions $\{p_i\}$. 
We call it a Helstrom family of ensembles if the Helstrom ratio $p = \frac{p_i}{\tilde{p}_i} $ attains the Helstrom bound: $ P_S = p$.  
\end{Def}
From equations \eqref{eq:keyineq}, an observable ${\bf E}$ satisfies $P_S({\bf E}) = p$ if 
$e_i(t_i) = 0$ for any $i=1,\ldots,N$. 
Then, it follows $p = P_S({\bf E}) \le P_S \le p$. 
Consequently, we have  
\begin{Prop}\label{Prop:SC}
A sufficient condition for a weak Helstrom family of ensembles $\{\tilde{p}_i,s_i; 1 - \tilde{p}_i, t_i\} \ (i=1,\ldots,N)$ to be Helstrom family is that there exists an observable ${\bf E}=\{e_i\}_{i=1}^N$ satisfying $e_i(t_i) = 0$ for all $i=1,\ldots,N$. In this case, the observable ${\bf E}$ gives an optimal measurement to discriminate $\{s_i\}$ with a prior distribution $\{p_i\}$.   
\end{Prop}
Two states $t_1,t_2 \in \SA$ are said to be distinguishable if there exists an observable ${\bf E}=\{e_1,e_2\}$ which discriminates $t_1$ and $t_2$ with certainty (for any prior distributions), or equivalently satisfy
\begin{equation}\label{eq:binarydiscrimination}
e_1(\sigma_1) = 1, e_1(\sigma_2) = 0 \ (\Leftrightarrow  \ e_2(\sigma_1) = 0, e_2(\sigma_2) = 1).
\end{equation} Therefore, as a corollary of Proposition \ref{Prop:SC} for $N=2$, we obtained the following theorem for a binary state discrimination ($N=2$). 
\begin{Thm}\label{thm:HE2}
Let $\{\tilde{p}_i,s_i; 1 - \tilde{p}_i, t_i\} \ (i=1,2)$ be a weak Helstrom family of ensembles for states $s_1,s_2 \in \SA$ and a binary probability distribution $p_1,p_2$ such that  $t_1$ and $t_2$ are distinguishable states. 
Then, $\{\tilde{p}_i,s_i; 1 - \tilde{p}_i, t_i\} \ (i=1,2)$ is a Helstrom family with the Helstrom ratio $p= P_S$. 
An optimal measurement to distinguish $s_1$ and $s_2$ is given by an observable to distinguish $t_1$ and $t_2$. 
\end{Thm}
{\bf Proof} The distinguishability of $t_1$ and $t_2$ satisfies the sufficient condition in Proposition \ref{Prop:SC}.  
\hfill{$\blacksquare$}

\begin{figure} 
\includegraphics[height=0.2 \textwidth]{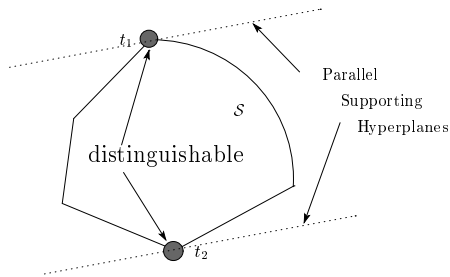}
\caption{Geometrical appearance of two distinguishable states $t_1,t_2$. }\label{fig:DS}  
\end{figure}

\begin{figure} 
\includegraphics[height=0.3 \textwidth]{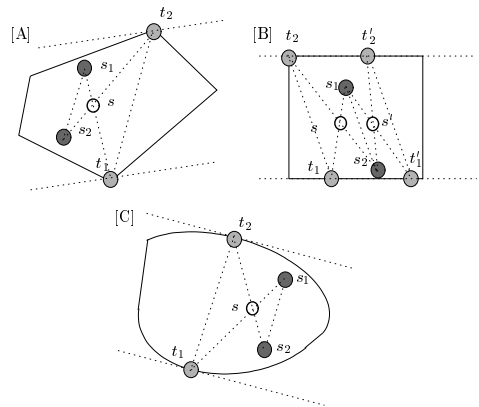}
\caption{[A] A typical Helstrom family of ensembles in $\R^2$; [B] a Helstrom family of ensembles is not unique; [C] A Helstrom family of ensembles exists for models $\SA$ with infinite numbers of extreme points.  }\label{fig:model}  
\end{figure}

Let us consider the case where $\SA$ is a subset of finite dimensional real 	vector space $V$.  
From condition \eqref{eq:binarydiscrimination}, geometrical meaning of two distinguishable states $t_1,t_2$ is that they are on the boundary of ${\cal S}$ which possess parallel supporting hyperplanes (See Fig.~\ref{fig:DS}). 
Here, a supporting hyperplane at a point $s \in \SA$ is a hyperplane $H \subset V$ such that $s \in H$ and $\SA$ is contained in one of the two closed half-spaces of the hyperplane \cite{ref:HP}. 
Indeed, if there exist two parallel supporting hyperplanes $H_1$ and $H_2$ at $t_1 \in \SA $ and $t_2 \in \SA$ respectively, one can construct an affine functional $f$ on $V$ such that $f(x) = 1$ on $x \in H_1 $ and $f(y) = 0$ for $y \in H_2$. 
Then, the restriction of $f$ to $\SA$ is an effect which distinguishes $t_1$ and $t_2$ with certainty since $\SA$ is contained between $H_1$ and $H_2$ and $f(t_1)=1,f(t_2)=0$.  
Then, to find a Helstrom family of ensembles given in Theorem \ref{thm:HE2} is nothing but a simple geometrical task. 
Here, we explain this in the uniform distribution cases: 
From the definition of a (weak) Helstrom family of ensembles and Theorem \ref{thm:HE2}, 
two ensembles $\{\tilde{p}_i,s_i; 1 - \tilde{p}_i, t_i\} \ (i=1,2)$ for a distinct stats $s_1,s_2 \in \SA$ with the uniform distribution $p_1=p_2=1/2$ are ensembles of a Helstrom family if $t_1,t_2$ are distinguishable and 
\begin{equation}\label{eq:HE2Eq}
s := q s_1 + (1-q) t_1 = q s_2 + (1-q) t_2, 
\end{equation}
with some $0 \le q:= \tilde{p}_1= \tilde{p}_2 \le 1$. 
From \eqref{eq:HE2Eq}, $s_1- s_2$ and $t_1 - t_2$ should be parallel, and therefore one easy way to find Helstrom family is as follows: search conjugate states $t_1$ and $t_2$ on the boundary of $\SA$ which are on a line parallel to $s_1-s_2$ such that there exist parallel supporting hyperplanes at $t_1$ and $t_2$. 
Then, the crossing point is a reference state $s$ while the ratio between $s_1-s$ ($s_2-s$) and $s-t_1$ ($s-t_2$) determines the Helstrom ratio $p= \frac{1}{Nq}$. In Fig.~\ref{fig:model}, Helstrom families for some models on $V = \R^2$ are illustrated. 

Now it is important to ask whether a Helstrom family of ensembles always exists for any general probabilistic theories or not.  
In this paper, we show a Helstrom family of ensembles for a binary state discrimination ($N=2$) always exist in generic cases for both classical and quantum systems. (For the existence in more general general probabilistic theories, see our forthcoming paper \cite{ref:ExistHE}.)  
Here, we mean by generic cases all the cases except for trivial cases where $P_S = \max[p_1,p_2]$ with a trivial measurement $u$, i.e., there are no improvement of guessing which exceeds the prior knowledge. 

First, let us consider a quantum system $\SA_{\mathrm{qu}}$. 
For distinct density operators $\rho_1,\rho_2$ with a prior probability distribution $p_1,p_2$, define an Hermitian operator $X := p_1 \rho_1 - p_2 \rho_2$. 
Let $X = \sum_i x_i P_i$ be the spectral decomposition of $X$. 
The positive and negative parts of $X$ are given by $X_+ := \sum_{i:x_i \ge 0} x_i P_i$ and $X_- := \sum_{i:x_i < 0} |x_i| P_i$ satisfying
\begin{equation}\label{eq:PN}
X = X_+ - X_-. 
\end{equation}
Note that $X_+,X_- \ge 0$, $X_+ X_- = 0$, and $||X_+||_1 - ||X_-||_1 = \tr X_+ - \tr X_- = p_1 - p_2$. 
$X_+$ or $X_-$ might be zero operator \footnote{Note that this happens even when $0 < p_1,p_2 < 1$. }, but in that case the optimization problem is nothing but a trivial case. 
Indeed, suppose that $X_- = 0$. 
Then, for any POVM element $E$, it follows $\tr E X = \tr E X_+ \le \tr \I X_+ = \tr X = p_1 - p_2$, and thus $P_S = p_1$ with a trivial measurement $\I$ from \eqref{eq:OS}. 
The similar argument shows that the case $X_+ =0 $ is again a trivial case with $P_S = p_2$. 
Therefore, for any generic case, we can assume $X_+,X_- \neq 0$, and this makes possible to define two density operators by 
\begin{equation}\label{eq:sigma}
\sigma_1 : = \frac{1}{||X_-||_1}X_-, \ \sigma_2 := \frac{1}{||X_+||_1}X_+. 
\end{equation} 
Notice that they are orthogonal and thus are distinguishable with certainty.   
It follows that $\sup_{0 \le E \le \I} \tr E X = \tr X_+ = ||X_+||_1$ where the maximum is established by the projection operator $P = \sum_{i;x_i \ge 0} P_i$. 
From \eqref{eq:OS}, we have 
\begin{equation}\label{eq:psq}
P_S^{(Q)} = p_2 + ||X_+||_1 = p_1 + ||X_-||_1.  
\end{equation}
Let $\tilde{p}_i = p_i / P^{(Q)}_S\ (i=1,2)$. It follows $0 < \tilde{p}_i \le 1$ from \eqref{eq:psq} and $\frac{p_1}{\tilde{p}_1}=\frac{p_2}{\tilde{p}_2}$ by definition. 
Finally, direct calculation using \eqref{eq:PN}, \eqref{eq:sigma} and \eqref{eq:psq} shows the equation \eqref{eq:HE}. 

Therefore, we have obtained \footnote{Although we explained in finite level quantum systems, it is straightforward to 
generalize to any quantum mechanical systems with Hilbert space with countably infinite dimension. (Notice that $X$ is a trace class operator on $\HA$ and thus has a discrete spectral decomposition.) }
\begin{Thm}\label{thm:QHE} 
For any quantum mechanical systems, a  Helstrom family of ensembles for a binary state discrimination exists for any generic cases. 
\end{Thm}
As any classical systems is embeddable into quantum systems (into diagonal elements with a fixed base), we have also 
\begin{Thm}\label{thm:CHE}
For any classical systems $\SA_{\mathrm{cl}}$, a Helstrom family of ensembles for a binary state discrimination exists for any generic cases. 
\end{Thm}
More concretely, for given distinct classical states $s_1 = (x_i)_{i=1}^d$, $s_2= (y_i)_{i=1}^d \in \SA_{\mathrm{cl}}$ ($x_i,y_i \ge 0, \sum_i x_i = \sum_i y_i = 1$) with a prior probability distribution $p_1,p_2$, one can define $
t_1 = \frac{1}{||X_-||_1}(-\min [X_i,0])_{i=1}^d$, $ t_2 = \frac{1}{||X_+||_1}(\max [X_i,$ $0])_{i=1}^d
$ where $X_i = p_1 x_i - p_2 y_i$, $||X_-||_1 = \sum_{i:X_i < 0} X_i$ and $||X_+||_1 = \sum_{i:X_i \ge 0} X_i$. 
Finally $\tilde{p}_i $ is given by $p_i / P_S $ $= 2 p_i / (1 + \sum_i |X_i|)$.  

In reference \cite{ref:Hwang}, a family of ensembles in Theorem \ref{thm:HE2} (and thus a Helstrom family of ensembles) has been used in $2$-level quantum systems for a binary state discrimination with a uniform prior distribution $p_0=p_1=1/2$.  
The purpose there was to reproduce Helstrom bound \eqref{eq:OS} in $2$-level quantum systems (with $p_0=p_1=1/2$) by resorting to (A) remote state preparation and (B) no-signaling condition \footnote{
{\bf [(A) Remote State Preparation]} Let $\{p_i;\rho_i\}$ and $\{q_j,\sigma_j\}$ be two ensembles in a quantum system ($\rho_i,\sigma_j \in \SA_{\mathrm{qu}}, \ p_i,q_j \ge 0, \sum_i p_i,\sum_j q_j = 1$) which satisfies $\sum_{i} p_i \rho_j = \sum_j q_j \sigma_j$. 
Then, there exist a Hilbert space $\cal K$, a state $\tau$ on $\HA\otimes {\cal K}$, and local measurements $M_1$ and $M_2$ on $\cal K$ such that the ensembles $\{p_i;\rho_i\}$ and $\{q_j,\sigma_j\}$ can be remotely prepared by measuring $M_1$ and $M_2$ under the state $\tau$. {\bf [(B) No-signaling condition]} Any information does not instantaneously transmit by local measurement. In quantum systems, it is well known that both (A) and (B) holds. 
}. 
Compared to their results, Theorem \ref{thm:QHE} shows that a Helstrom family of ensembles exists not only in $2$-level systems with uniform distributions but also in any quantum systems for generic cases. 
Moreover, Theorem \ref{thm:HE2} implies that a logical connection with an optimal state discrimination has already appears through the existence of a Helstrom family of ensembles, resort to neither (A) nor (B); and indeed this appears in any general probabilistic theories, not only in quantum systems. 
Of course, this is consistent with the results in \cite{ref:Hwang} and our result can be interpreted as a generalization of the results in \cite{ref:Hwang} to any dimensional quantum mechanical systems for any $N$ states discrimination. 

\section{Examples}\label{sec:ex}

In this section, we illustrate our method in quantum $2$-level systems (qubit), and also in a simple toy model which is neither classical nor quantum. 

\subsection{Quantum $2$-level systems}

As is well known, any density operator $\rho$ for qubit is represented by the Bloch vector ${\bm b} \in D^3:= \{{\bm b} \in \R^3 \ | \ ||{\bm b}|| \le 1\}$ through the map ${\bm b} \mapsto \rho({\bm b}) = \frac{1}{2}(\I + \sum_{i=1}^3 b_i \sigma_i)$, where $\sigma_i \ (i=1,2,3)$ are Pauli Matrices. 
Notice that the trace distance between density operators coincides with the Euclid distance in $\R^3$ between the corresponding Bloch vectors:  $||\rho({\bm b}_1) - \rho({\bm b}_2)||_1 = ||{\bm b}_1-{\bm b}_2||$. 

\bigskip 

\begin{figure} 
\includegraphics[height=0.3 \textwidth]{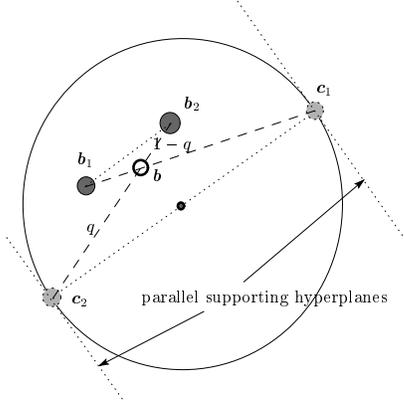}
\caption{$2$-dimensional section of the Bloch Ball. }\label{fig:Bloch}
\end{figure}

[Examples 3: Binary state discrimination] Let us consider a state discrimination between $\rho({\bm b}_1)$ and $\rho({\bm b}_2) $ with a uniform distribution. 
Following a geometrical view of a Helstrom family of ensembles in Theorem \ref{thm:HE2}, one can find it in the following manner:  
In order that states ${\bm c}_1 \in D^3$ and ${\bm c}_2 \in D^3$ have parallel supporting hyperplanes so that they are distinguishable, 
they should be on the Bloch sphere (pure states) in opposite direction \footnote{Indeed, this is equivalent to the orthogonality between $\rho({\bm c}_1)$ and $\rho({\bm c}_2)$. }. 
Moreover, the line ${\bm c}_1 -{\bm c}_2$ has to be parallel to ${\bm b}_1 - {\bm b}_2$ from condition \eqref{eq:HE2Eq}. 
Then, ${\bm c}_1$ and ${\bm c}_2$ are uniquely determined by points on the intersection of the Bloch ball and the hyperplane determined by ${\bm b}_1-{\bm b}_2$ and the origin (See Fig.~\ref{fig:Bloch}). 
Then, it is an elementary geometric problem to obtain the ratio: $q = \frac{2}{2+||{\bm b}_1 - {\bm b}_2||}$. 
Since the Helstrom ratio is given by $p = p_i/\tilde{p}_i = 1/2q$, this reproduces a Helstrom bound $P_S = \frac{1}{2} (1 + \frac{1}{2}||{\bm b}_1-{\bm b}_2||)$ by use of Theorem \ref{thm:HE2}. 
Indeed, from \eqref{eq:OS}, the optimal success probability to discriminate two distinct $\rho_1$ and $\rho_2$ with a uniform prior distribution is 
\begin{eqnarray}\label{eq:OS2}
P^{(Q)}_{S} = \frac{1}{2} (1 + \frac{1}{2}||\rho_1 - \rho_2||_1).  
\end{eqnarray}
(Recall that $||\rho({\bm b}_1) - \rho({\bm b}_2)||_1 = ||{\bm b}_1-{\bm b}_2||$). 

\bigskip 

\begin{figure}
\includegraphics[height=0.25 \textwidth]{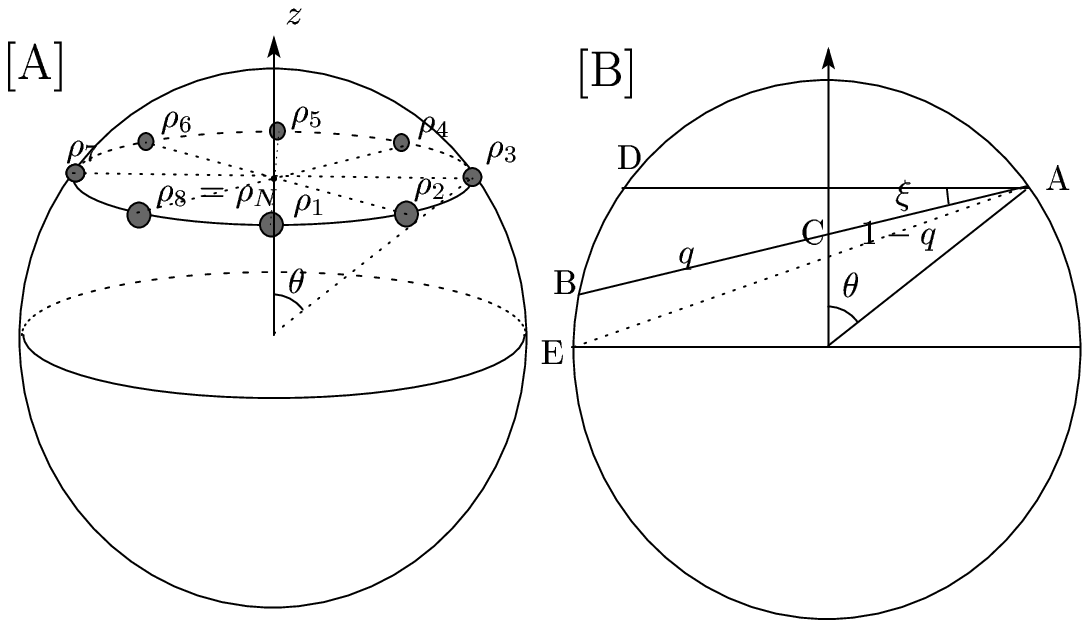}
\includegraphics[height=0.25 \textwidth]{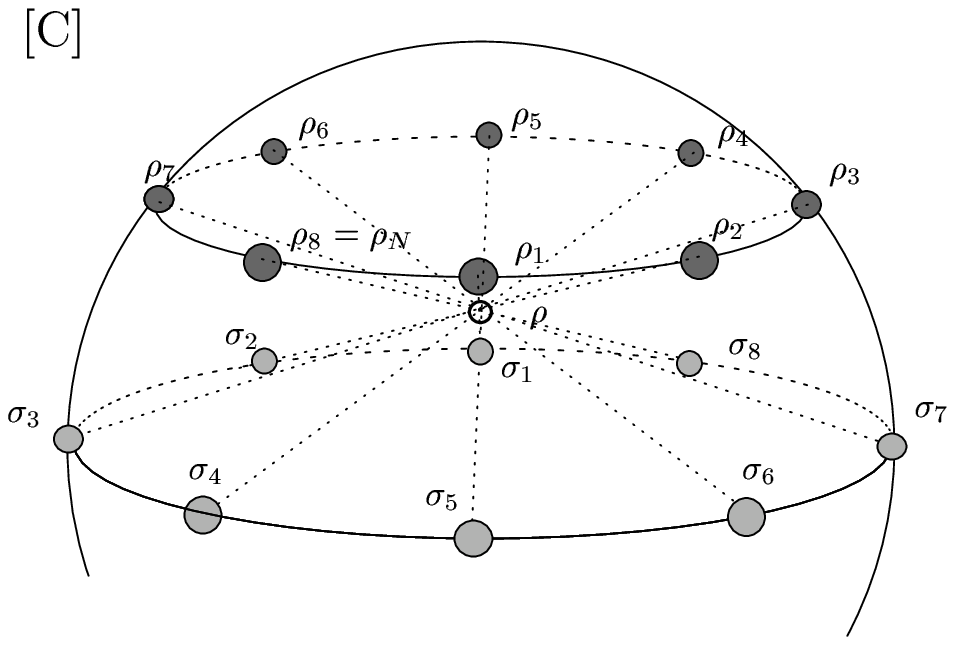}
\caption{[A] Symmetric quantum states $\rho_1,\ldots,\rho_8$ in the Bloch ball;  
[B] $2$-dimensional section where points A, B, C are $\rho_j$, $\sigma_j$ and $\rho$, respectively;
($\overline{AC} = \frac{\sin \theta}{ \cos \xi}$, $\overline{AB} = 2 \sin (\theta + \xi)$, and thus 
$q = 1- \frac{\overline{AC}}{\overline{AB}} = 1 - \frac{\sin \theta}{ 2 \sin (\theta + \xi)\cos \xi}$.);  
[C] Helstrom family of ensembles with conjugate states $\sigma_j$ and the reference state $\rho$. }\label{fig:BlochN}
\end{figure}

[Examples 4: $N$-numbers of symmetric state discrimination] In quantum systems, discrimination of $N$ numbers of state is much more difficult problem than binary cases. 
For symmetric quantum (pure) states $\{\rho_j = \ketbra{\psi_j}{\psi_j}\}_{j=1}^N$ with uniform distribution $p_i = 1/N$, where state vectors are given by $\ket{\psi_j} = V^{j-1} \ket{\psi}$ with a normalized vector $\ket{\psi}$ and a unitary operator satisfying $V^N = \exp(i \chi )\I \ (\chi \in \R)$, Ban et al. \cite{ref:ban} obtained the optimal success probability:  
$$
P_S^{(Q)} = |\bra{\psi}\Phi \ket{\psi}|^2,
$$  
where $\Phi : = \sum_{j=1}^N \ketbra{\psi_j}{\psi_j}$. 
As a typical example, let us consider $N$ symmetric states in $2$-level systems (as illustrated in Fig.~\ref{fig:BlochN} [A] for the case $N=8$). 
Let $V:= \exp(-i  \frac{\pi}{N} \sigma_3)$ be a unitary operator which rotates Bloch vectors by the angle $2\pi/N$ around on the $z$ axis ($V^N = -\I$); and let $\ket{\psi} := \cos(\frac{\theta}{2})\ket{0} + \sin(\frac{\theta}{2})\ket{1}$ where $\ket{0},\ket{1}$ the eigenvectors of $\sigma_3$ with eigenvalues $1,-1$. The corresponding Bloch vector to $\ket{\psi}$ is ${\bm b} = (\sin \theta,0,\cos \theta)$.
Then, it follows 
\begin{equation}
\ket{\psi_j} := V^{j-1} \ket{\psi
}= \cos(\frac{\theta}{2})\ket{0} + \sin (\frac{\theta}{2}) e^{i \frac{2 \pi (j-1)}{N}}\ket{1}, 
\end{equation} 
for $j=1,\ldots,N$, with the corresponding Bloch vectors ${\bm b^{(j)}} = (\sin \theta \cos \frac{2 \pi (j-1)}{N} ,\sin \theta \sin \frac{2 \pi (j-1)}{N},\cos \theta)$.  
It is easy to show $\Phi = \frac{N}{2}(\I + \cos \theta \sigma_3)$ \footnote{For instance, one can make use of the Bloch vector representation. $\Phi := \sum_{j=1}^N \ketbra{\psi_j}{\psi_j}  = \sum_{j=1}^N \frac{1}{2}(\I + \sum_{k=1}^3 b^{(j)}_k \sigma_k) = \frac{N}{2}(\I + \cos \theta \sigma_3)$, since $\sum_j b^{(j)}_1 \sigma_1 = \sum_j b^{(j)}_1 \sigma_1 $ vanish from the rotational symmetry around $z$-axis. }, and the optimal success probability is 
\begin{equation}\label{eq:PSsym}
P_S^{(Q)} = \frac{1}{N}(1+ \sin \theta). 
\end{equation}
In the following, we apply our method and show that there exists a Helstrom family of ensemble for this problem with any $N$ and thus reproduce the success probability \eqref{eq:PSsym}.   
(In the following, we identity the density operator $\rho_j$, the state vector $\psi_j$, and its Bloch vector ${\bm b_j}$.  

First, from the symmetry and geometrical view point of a weak Helstrom family of ensembles, it is clear that a weak Helstrom family for $\{\rho_j =\ketbra{\psi_j}{\psi_j}\}_{j=1}^N$ and $p_i = 1/N$ can be constructed as follows: 
In the Bloch ball, make lines from each $\rho_j$ to a point on the $z$-axis, say point $C$,  and extend the lines until they arrives at the Bloch sphere, and let conjugate states $\sigma_j$ be each end-points of the lines from $\rho_j$.    
Fig.~\ref{fig:BlochN} [B] shows one of the $2$-dimensional sections of the Bloch ball where the points A and B are the corresponding $\rho_j$ and $\sigma_j$.  
Then, we have obtained a weak Helstrom family of ensembles $\{q_\xi,\rho_j;1-q_\xi, \sigma_j\}$ where $q_\xi$ is a ratio $\frac{\overline{CB}}{\overline{AB}}$, where we explicitly write the dependence on the angle $\xi = \angle DAB$, so that the reference state $\rho$ is the point C: 
$$
\rho = q_\xi \rho_j + (1-q_\xi) \sigma_j \ (j=1,\ldots,N).     
$$ 
Note that we have a bound $P_S \le p= \frac{1}{N q_\xi}$ from Proposition \ref{Prop:BDD}. 
Therefore, in order to obtain a tighter bound of $P_S$, we would like to find a weak Helstrom family with larger $q_\xi$ as much as possible. 
It is again a simple geometric problem to obtain $q_\xi = 1 - \frac{\sin \theta}{\sin (\theta + 2 \xi ) + \sin \theta}$ (see the caption of Fig.~\ref{fig:BlochN} [B]), which takes the maximum $q_{\xi_M} = \frac{1}{1 + \sin \theta} $ at $\xi_M = \frac{\pi}{4} - \frac{\theta}{2} \ (= \angle  DAE)$ (See Fig.~\ref{fig:BlochN} [C]). 
This attains the tight bound \eqref{eq:PSsym}, and thus we have demonstrated that our method reproduces the optimal success probability. 
Indeed, we can show that this weak Helstrom family of ensembles is a Helstrom family from Proposition~\ref{Prop:SC}: 
note that $\sigma_j = \ketbra{\phi_j}{\phi_j}$ at $\xi_M$ is  
\begin{equation}
\ket{\phi_j }= \cos(\frac{\pi}{4})\ket{0} + \sin (\frac{\pi}{4}) e^{i (\frac{2 \pi (j-1)}{N} + \pi)}\ket{1}.
\end{equation}
Let $\ket{\chi_j}:= \cos(\frac{\pi}{4})\ket{0} + \sin (\frac{\pi}{4}) e^{i \frac{2 \pi (j-1)}{N}}\ket{1}$ which is orthogonal to $\ket{\phi_j }$ for all $j=1,\ldots,N$. 
Then, it follows $\sum_{j=1}^N \ketbra{\chi_i}{\chi_i} = \frac{N}{2} \I$ and thus $\{E_i := \frac{2}{N} \ketbra{\chi_i}{\chi_i}\}$ is a POVM which satisfies the condition $\tr E_i \sigma_i = 0$ in Proposition~\ref{Prop:SC}. 
Consequently, we have found a Helstrom family of ensembles and thus obtained the Helstrom bound.    

\subsection{Probabilistic Model with square-state space} 

\begin{figure}
\includegraphics[height=0.23 \textwidth]{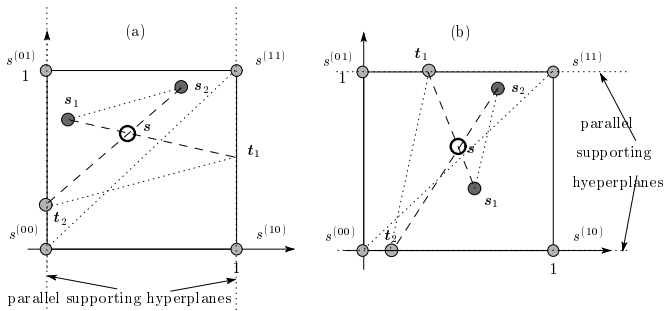}
\caption{Probabilistic model with square-state space.}\label{fig.sq}
\end{figure}

As an example which is not neither classical nor quantum systems, let us consider a general probabilistic model with square-state space 
$\SA_{sq} : = \{(x_1,x_2)\in \R^2 \ | \ 0 \le x_i \le 1 (i=1,2)\}$ (Fig.~\ref{fig.sq}).  
This can be considered as a simplest nontrivial model which is neither classical nor quantum systems.  
It should be noticed that this is not just a toy model and one can show that this probabilistic model can be physically realized from a classical system under a certain restriction of measurements \cite{ref:HolevoSM,ref:KIMNI}.
It is obvious that $\SA_{sq}$ is a compact convex subset in $\R^2$ with 4 numbers of pure states:
\begin{equation}\label{eq:purestates}
{\bm s}^{(00)} = (0,0), {\bm s}^{(01)} =(0,1),{\bm s}^{(10)} =(1,0), {\bm s}^{(11)} =(1,1). 
\end{equation}

[Example 5: Binary state discrimination] Let us consider a binary state discrimination problem for two distinct states ${\bm s_1}=(x_1,y_1),{\bm s_2}=(x_2,y_2) \in \SA_{sq}$ with uniform distribution. Without loss of generality, let $x_1 \le x_2$. 
There are two cases; (a) $\angle ({\bm s_2}-{\bm s_1},{\bm s}^{(10)}-{\bm s}^{(00)}) \le \pi/4$ or (b) $\angle ({\bm s_2}-{\bm s_1},{\bm s}^{(10)}-{\bm s}^{(00)}) \ge \pi/4$, where $\angle ({\bm a},{\bm b}) := \arccos(\frac{{\bm a}\cdot{\bm b}}{\sqrt{{\bm a}\cdot{\bm a}}\sqrt{{\bm b}\cdot{\bm b}}})$ is the angle between two vectors ${\bm a}$ and ${\bm b}$. 
In case (a), clearly there exist conjugate states ${\bm t}_1$ and ${\bm t}_2$ on line ${\bm s}^{(11)}-{\bm s}^{(10)}$ and line ${\bm s}^{(01)}-{\bm s}^{(00)}$, respectively such that ${\bm t}_1 - {\bm t}_2$ are parallel to ${\bm s}_1 - {\bm s}_2$.  
Since there exists parallel supporting hyperplanes on $t_1$ and $t_2$ (see Fig.~\ref{fig.sq} (a)), we have a Helstrom family from Theorem \ref{thm:HE2}. 
Then, it is an elementary calculation to find $q=\frac{1}{1+|x_2-x_1|}$, and hence the optimal success probability is $P_S = \frac{1}{2}(1+|x_2-x_1|)$; Similarly in case (b), we have a Helstrom family and the optimal success probability is given by 
$P_S = \frac{1}{2}(1+|y_2-y_1|)$  (see Fig.~\ref{fig.sq} (b)). 
\bigskip 

[Example 5: state discrimination of pure states]  Since $\SA_{sq}$ is not a simplex, and thus not a classical system, four pure states \eqref{eq:purestates} cannot be discriminated in a single measurement. 
Let us obtain the optimal success probability to distinguish all the pure states with uniform distribution.
From a geometrical consideration, one has to find as large polygon as possible in $\SA_{sq}$ which is reverse homothetic to $\mathrm{conv}_{i,j=0,1}[{\bm s}^{(ij)}] = \SA_{sq}$. 
Clearly, it is $\SA_{sq}$ itself, with the similarity point at the center of $\SA_{sq}$. 
More precisely, one can choose conjugate states ${\bm t}^{(ij)} = {\bm s}^{(i\oplus 1,j\oplus 1)}$ where $\oplus$ denotes the exclusive OR, and $q = 1/2$. 
Therefore, we obtained a weak Helstrom family with the Helstrom ratio $p = \frac{1}{4q}= 1/2$. 
It turns out that this weak Helstrom family is a Helstrom family, and thus we obtain $P_S = 1/2$ to discriminate all pure states in this system. 
Indeed, it is easy to see that affine functionals $e^{(ij)} \ (i,j=0,1)$ on $\SA_{sq}$ defined by $e^{(ij)}({\bm t}^{ij}) = 0, \ e^{(ij)}({\bm t}^{i\oplus 1, j\oplus 1}) = 1/2$ (and hence satisfying $e^{(ij)}({\bm t}^{i\oplus 1, j})=e^{(ij)}({\bm t}^{i, j\oplus 1}) = 1/4$) for any $i,j=0,1$ forms a $4$-valued observable $\{e^{(ij)}\}$ on $\SA_{sq}$.  
This satisfies the sufficient condition in Proposition \ref{Prop:SC}.    
 
\section{Conclusion}\label{ref:2}

In this paper, we introduced a notion of a (weak) Helstrom family of ensembles in general probabilistic theories and showed the close relation with state discrimination problems. 
Basically,  Helstrom family can be searched by means of geometry, and once we have the family, or at least a nontrivial weak family, the optimal success probability, or a bound of it, is automatically obtained from the Helstrom ratio. 
In binary state discriminations, a weak Helstrom family of ensembles with distinguishable conjugate states is shown to be a Helstrom family which has again a simple geometrical interpretation. 
We illustrated our method in $2$-level quantum systems and reproduced the Helstrom bound \eqref{eq:OS2} for binary state discrimination and symmetric quantum states \eqref{eq:PSsym}.
As an nontrivial general probabilistic theories, a probabilistic model with square-state space is investigated and binary state discrimination and pure states discrimination are established using our method.  
In this paper, we showed the existences of Helstrom families of ensembles analytically in both classical and quantum theory in any generic cases in binary state discriminations. 
For the more general models, it will be investigated in our forthcoming paper \cite{ref:ExistHE}. 
There, we also clarify the relation between our method and linear programming problem.

 {\bf Acknowledgment}

We would like to thank Dr. Imafuku, Dr. Nuida and Dr. Hagiwara for their fruitful comments.

\end{document}